\documentclass[prd]{revtex4}% Physical Review D
\usepackage{array}
\usepackage{booktabs}
\usepackage{tabu}
\usepackage{dcolumn}
\usepackage{amsmath}
\usepackage{amsfonts}
\usepackage{amssymb}
\usepackage{graphicx}
\usepackage{subfigure}
\usepackage{graphicx}% Include figure files
\usepackage{dcolumn}% Align table columns on decimal point
\usepackage{bm}% bold math

\begin{document}

\title{The effect of de Sitter like background on increasing the zero point budget of dark energy}
  \author{Haidar Sheikhahmadi$^{a,b}$}
   \email{h.sh.ahmadi@iasbs.ac.ir/ gmail.com}
    \author{Ali Aghamohammadi$^{c}$}
    \email{a.aghamohamadi@iausdj.ac.ir}
      \author{Khaled Saaidi$^{b}$}
       \email{ksaaidi@uok.ac.ir}
\affiliation{$^a$Institute for Advance Studies in Basic Sciences (IASBS), Gava Zang, Zanjan 45137-66731, Iran\\
$^b$Department of Physics, Faculty of Science, University of Kurdistan,  Sanandaj, Iran\\
$^c$Sanandaj Branch, Islamic Azad University, Sanandaj, Iran}
\date{\today}% It is always \today, today,

\def\be{\begin{equation}}
  \def\ee{\end{equation}}
\def\bea{\begin{eqnarray}}
\def\eea{\end{eqnarray}}
\def\f{\frac}
\def\n{\nonumber}
\def\l{\label}
\def\p{\phi}
\def\o{\over}
\def\R{\rho}
\def\pa{\partial}
\def\om{\omega}
\def\na{\nabla}
\def\P{\Phi}
\def\g{\gamma}
%\nofiles

%=============================================================%
%=============================================================%
%============== Abstract =======================================%
%=============================================================%
%=============================================================%
\begin{abstract}
\section*{Abstract}
During this work, using subtraction renormalization mechanism, zero point quantum fluctuations for  bosonic scalar fields  in a de-Sitter like background are investigated. By virtue of the observed value for spectral index, $n_s(k)$, for massive scalar field the best value for the first slow roll parameter, $\epsilon$, is achieved. In addition the energy density of vacuum quantum fluctuations for massless scalar field is obtained. The effects of these fluctuations on other components of the Universe are studied. By solving the conservation equation, for some different examples, the energy density for different components of the Universe are obtained. In  the case which, all components of the Universe are in an interaction, the different dissipation functions, $\tilde{Q}_{i}$, are considered. The time evolution of  ${\rho_{DE}(z)}/{\rho_{cri}(z)}$ shows that $\tilde{Q}=3 \gamma H(t) \rho_{m}$ has best  agreement in comparison to observational data including  CMB, BAO and SNeIa data set.
\end{abstract}
\pacs{.....}
\maketitle
%%%%%%%%%%%%%%%%%%%%%%%%%%%%%%%%%%%%%%%%%%%%%%%%%%%%%%%%%%%%%
%%%%%%%%%%%%%%%%%%%%%%%%%%%%%%%%%%%%%%%%%%%%%%%%%%%%%%%%%%%%%
%============  Sec.I (Introduction)  =======================================
%%%%%%%%%%%%%%%%%%%%%%%%%%%%%%%%%%%%%%%%%%%%%%%%%%%%%%%%%%%%%
%%%%%%%%%%%%%%%%%%%%%%%%%%%%%%%%%%%%%%%%%%%%%%%%%%%%%%%%%%%%%
\section{Introduction}
\label{sec1}
\indent
For a de-Sitter like background, to investigate the effects of the quantum fluctuations on the energy budget of the Universe  both massive and massless scalar fields  are considered. Because of the appearance of the first slow roll parameter, the relation between power spectral index and slow roll parameters leads to a away to connect theoretical results with observations. In fact one can estimate the best value of the first slow roll parameter based on observations, for instance Planck 2013 {\cite{Planck}}. From this perspective, for zero point quantum fluctuations both energy density and pressure are calculated.  Interestingly it is observed that, the contribution of zero point energy is increased against  the results for the normal de-Sitter ones {\cite{11, 12}}. The importance of this note is that whereas the zero point contribution of dark energy potentially is detectable, therefore the possibility of dark energy detection is increased. Also it should be stressed, due to time dependency of Hubble parameter which is appeared in energy density of zero point quantum fluctuations, energy can transfer between different components of the Universe. By considering this concept one can propose different manners to investigate interaction between zero point fluctuations and other sectors. In a general case which all components are in an interaction, three different dissipation function $\tilde{Q}$ are considered. For this special case time evolution of  ${\rho_{DE}(z)}/{\rho_{cri}(z)}$ shows that $\tilde{Q}=3 \gamma H(t) \rho_{m}$ is in best agreement in comparison to observations including  CMB, BAO and SNeIa data set. Furthermore, energy density of matter, $\rho_{m}$, has some deviations in comparison to the result of ordinary de-Sitter one. In fact in $\rho_{m}$, beside $a^{-3}$, some extra terms are appeared, which these new terms could be interpreted as new source for cold dark matter risen from interaction between matter and  quantum fluctuations. In fact as it was discussed in {\cite{12}} zero point quantum fluctuations could be considered as sub-dark energy, these extra terms in question can be proposed as sub-dark matter. The scheme of this paper is as follows:\\
 In Sec.$II$ the general framework of this work including the mathematical calculations are discussed. In Sec.$III$ the cosmological role for zero point energy density is investigated, and the results of this work are compared with previous works. In Sec.$IV$ to estimate the amount of sub-dark energy and also sub-dark matter, the bounds which risen from time evolution of dark energy are considered.  And at last, we have conclusion.
%%%%%%%%%%%%%%%%%%%%%%%%%%%%%
%%%\section{General framework}%%%%%
%%%%%%%%%%%%%%%%%%%%%%%%%%%%
\section{Massive Scalar Field and Slow Roll Parameters}
To study the effect of zero point quantum fluctuations let's  consider a real minimally coupled bosonic scalar field $\Phi$ in a semiclassical general relativity mechanism {\cite{Cas}}. In such scenario the  geometry is not quantized but the energy momentum tensor related to the scalar field is calculated by means of the vacuum expectation value concept. To begin we consider the action
\begin{equation}\label {1-3}
S=S_m+\int d^4x\frac{\sqrt{-g}}{\kappa^2} \left(R-2\Lambda+\kappa^2 \mathcal{L}_{\Phi} \right),
\end{equation}
where  $S_m$ is the matter action, $g$ is the  determinant of the metric, $R$ is the Ricci scalar, $\Lambda$ is the
Einstein's cosmological constant and $\mathcal{L}_{\Phi}$ is defined as
\begin{equation}\label {2-3}
\mathcal{L}_{\Phi}=-\frac{1}{2}g^{\mu\nu}(\nabla_{\mu}\Phi\nabla_{\nu}\Phi)-V(\Phi),
\end{equation}
where $V(\Phi)$ is the potential of the model. Variation the action (\ref{1-3}) with respect to metric yields
\begin{equation}\label {3-3}
R_{\mu\nu}-\frac{1}{2}g_{\mu\nu}R+g_{\mu\nu}\Lambda=\kappa^2(T^{M}_{\mu\nu}+<0\mid T^{\Phi}_{\mu\nu}\mid0>_{ren}),
\end{equation}
here   $T^{m}_{\mu\nu}= ({-2}/{\sqrt{-g}})\times({\delta S_m}/{\delta g^{\mu\nu}})$ and
\begin{equation}\label {3-33}
T^{\Phi}_{\mu\nu}=\partial_{\mu}\Phi\partial_{\nu}\Phi-\frac{1}{2}g_{\mu\nu}\big(\partial_{\alpha}\Phi\partial^{\alpha}\Phi+2V(\Phi)\big).
\end{equation}
Using above equation, the $00$ and $ii$ components of energy-momentum tensor reads
\begin{equation}\label {4-33}
T_{0}^{\Phi 0}=-\frac{1}{2}\left[\dot{\Phi}^{2}+\frac{\Phi_{,i}^{2}}{a^{2}(t)}+2V(\Phi)\right],\nonumber
\end{equation}
\begin{equation}\label {5-33}
\frac{1}{3}T_{i}^{\Phi i}=\frac{1}{2}\left[\dot{\Phi}^{2}-\frac{1}{3}\frac{\Phi_{,i}^{2}}{a^{2}(t)}-2V(\Phi)\right].
\end{equation}
It is also obvious that  variation the Eq.(\ref{1-3}) with respect to $\Phi$  yields
\begin{equation}\label {6-33}
\Box\Phi-V_{,\Phi}(\Phi)=0,
\end{equation}
where $\Box$ is d'Alembert operator.  As it mentioned above in semiclassical approach one can quantize the scalar field and therefore the vacuum expectation value of energy density and pressure could be obtained. For this purpose, the quantized scalar field is defined as
\begin{equation}\label {2}
\Phi(t, \mathbf{x})=\int \frac{d^3k}{(2\pi)^{3}\sqrt{2k}}\left[a_\mathbf{k}\phi(t)e^{-j\mathbf{k}.\mathbf{x}}+
a^{\dag}_\mathbf{k}\phi^{\ast}(t)e^{j\mathbf{k}.\mathbf{x}}\right],
\end{equation}
where $\phi(t)$ is a function which should be determined, $a_\mathbf{k}$ ($a^{\dag}_\mathbf{k}$) is annihilation (creation) operators and $\sqrt{-1}=j$  {\cite{13, 13-aa}}. Now introducing Eq.(\ref{2}) to (\ref{6-33}), it expresses
\begin{equation}\label {7-33}
\ddot{\phi}+3H(t)\dot{\phi}+\frac{k^2}{a^2}\phi+V_{,\phi}=0,
\end{equation}
where  $H=\dot{a}/a $, and $a(t)$  is the scale factor of the Universe, over-dot denotes derivation with respect to the cosmic time and the ${k^2\phi}/{a^2}$ term is appeared due to scalar field's spatial dependency. For more convenient one can use conformal time $\eta=\int dt/a(t)$, and therefore Eq.(\ref{7-33}) could be rewritten as
\begin{equation}\label {4}
\phi^{\prime\prime}+2\mathcal{H} \phi^{\prime}+ k^{2} \phi+a^2\frac{V^{\prime}(\phi)}{\phi^{\prime}}=0,
\end{equation}
in which over-prime indicates derivation with respect to $\eta$, and  $\mathcal{H}=a^{\prime}/a=\dot{a}$ is the conformal, comoving, Hubble parameter. Now by defining $\chi:=\phi/a$, using a power law potential, $V(\phi)=m^2\phi^2/2$, relation (\ref{4}) reads
\begin{equation}\label {5-3}
\chi^{\prime\prime}+ \big(k^{2}+m^2a^2{-\frac{a^{\prime\prime}}{a}}\big) \chi=0.
\end{equation}
It should be stressed for in question de-Sitter like background, scale factor is taken as $a(t)=t^{1/\epsilon}$, where $\epsilon=-\dot{H}/H^{2}\ll 1$ is the first slow roll parameter \cite{{14, 14-a}}. Using definitions of scale factor (i.e.  $a(\eta)\simeq-(1+\epsilon)/H\eta$) and second slow roll parameter, $\tau=m^2/3H^2$, one can attain
\begin{equation}\label {6-3}
-m^2a^2+{\frac{a^{\prime\prime}}{a}}\simeq\frac{1}{\eta^2}\Big(2+3(\epsilon-\tau)\Big).
\end{equation}
By introducing $\tilde{\nu}:=(\epsilon-\tau)+3/2$ and keep only up to first orders of the first and second slow roll parameters  Eq.(\ref{5-3}) could be rewritten as
\begin{equation}\label {7-3}
\chi^{\prime\prime}+ \big(k^{2}-{\frac{\tilde{\nu}^{2}-\frac{1}{4}}{\eta^2}}\big) \chi=0,
\end{equation}
where $\tilde{\nu}^{2}\simeq3(\epsilon-\tau)+9/4$. Solving above Bessel like equation, the magnitude of $\phi$ can be achieved as
\begin{equation}\label {8-3}
|\phi|\simeq\frac{H(1-\epsilon)}{k}\Big(\frac{k}{aH(1-\epsilon)}\Big)^{\frac{3}{2}-\tilde{\nu}}.
\end{equation}
To estimate the best value of $\epsilon$, one can use the power spectrum concept for instance in light of Planck $2013$ {\cite{Planck}}. It is obvious that by considering the Fourier transformation for an arbitrary function as
\begin{equation}\label {9-3}
\widetilde{\Phi}_{\mathbf{k}}(t, \mathbf{r})=\int\frac{d^3k}{(2\pi)^{3}\sqrt{2k}}e^{j \mathbf{k}.\mathbf{r}}\widetilde{\phi}_\mathbf{k}(t),
\end{equation}
where $\mathbf{k}$ is comoving momentum and $\mathbf{r}$ is spatial vector the power spectrum can be defined as
\begin{equation}\label {10-3}
<\widetilde{\phi}_{\mathbf{k}}(t) \widetilde{\phi}^{\ast}_{\mathbf{k}^{\prime}}(t)>=\frac{2\pi^{2}}{k^2}\rho_{S}(2\pi)^{3}\delta{^3}(\mathbf{k}-\mathbf{k}^{\prime}),
\end{equation}
where $\rho_{S}$ is power spectrum in question and $<\widetilde{\phi}_\mathbf{k}(t) \widetilde{\phi}^{\ast}_{\mathbf{k}^{\prime}}(t)>$ indicates the mean square value of $\widetilde{\phi}_\mathbf{k}(t)$. By combining Eqs.(\ref{2}), (\ref{9-3}) and (\ref{10-3}) the power spectrum can be achieved as
\begin{equation}\label {11-3}
\rho_{S}=\frac{k^{2}}{(2\pi)^{2}}|\phi(t)|^{2}.
\end{equation}
To achieve this result the relation between annihilation and creation operators reads $[a_\mathbf{k}, a^{\dag}_{\mathbf{k}^{\prime}}]=(2\pi)^{3}\delta{^3}(\mathbf{k}-\mathbf{k}^{\prime})$. At last if one consider the relation (\ref{8-3}), the power spectrum could be rewritten as
\begin{equation}\label {12-3}
\rho_{S}=(1-2\epsilon)H^{2}\Big(\frac{k}{aH(1-\epsilon)}\Big)^{{3}-2\tilde{\nu}}.
\end{equation}
Also it is notable, another important quantity which plays a crucial role in inflationary investigations is spectral index which is defined as
\begin{equation}\label {13-3}
n_S-1=\frac{d\ln{\rho_S}}{d\ln k}.
\end{equation}
Substituting Eq.(\ref{12-3}) into above equation, one has
\begin{equation}\label {14-3}
n_S-1=(1-2\epsilon)({{3}-2\tilde{\nu}}),
\end{equation}
and at last after some manipulations the spectral index  is obtained as $n_S-1=-2\epsilon+2\tau$.
By means of observed value of spectral index, $n_s<0.9675$, risen from Planck $2013$  {\cite{Planck}}, the best value of the first slow roll parameter is $\epsilon \simeq 0.02$.
%=====================================================
%==============  Typical example: Massless Scalar Field ==================================
%=======================================================
\subsection{Typical example: Massless Scalar Field}
In this case we want to consider massless scalar field. Although this case is an ideal example, but because of simplicity and also good estimations for physical results attracts more attention. Therefore if in action (\ref{1-3}), one assumes $V(\Phi)=0$, solving Eq.(\ref{4}), for positive modes, is attained as
\begin{equation}\label {5}
\phi(\eta)=\frac{1}{a(\eta)}\left(1-\frac{i\xi}{k\eta}\right)e^{(-ik\eta)},
\end{equation}
where $\xi=({2+3\epsilon})/2$. In this case, as massive ones, the first slow roll parameter is considered only up to the first order. For this typical case we want to estimate the zero point quantum fluctuation contribution in the energy budget of the Universe. To begin we have to calculate the vacuum expectation value  of the scalar field energy-momentum tensor. By means of  Eqs.(\ref{5-33}) and (\ref{5}), one has
\begin{equation}\label {---}
\rho_{vac}=<0|T^{\phi}_{00}|0>=\frac{1}{2}\int_{0}^{a(t)\Lambda_{c}}\frac{d^{3}k}{(2\pi)^{3}2k}
\left[|\dot{\phi}|^{2}+\frac{k^{2}}{a^{2}(t)}|\phi|^{2}\right],\nonumber
\end{equation}
\begin{equation}\label{8}
P_{vac}=\frac{1}{3}\Sigma_{i}<0|T_{i}^{\phi i}|0> =
\frac{1}{2}\int_{0}^{a(t)\Lambda_{c}}\frac{d^{3}k}{(2\pi)^{3}2k}
\left[|\dot{\phi}|^{2}-\frac{k^{2}}{3a^{2}(t)}|\phi|^{2}\right].
\end{equation}
 where indicate the energy density and pressure of the vacuum quantum fluctuations respectively. According to quantum field theory the cutoff $\Lambda_{c}$, should be considered greater than physical momenta $k/a(t)$. By considering Eq.(\ref{5}) and definition of scale factor for a de Sitter like background, one has
\begin{equation}
\left(|\dot{\phi}|^{2}+\frac{k^{2}}{a^{2}(t)}|{\phi}|^{2}\right)=\frac{1}{a^{4}}\left[2k^{2}+
{a^{2}H^{2}(1+9\epsilon)}\right]+\mathcal{O}(\epsilon^{2}),\nonumber
\end{equation}
\begin{equation}\label{11}
\left(|\dot{\phi}|^{2}-\frac{k^{2}}{3a^{2}(t)}|\phi|^{2}\right)=\frac{2}{3a^{4}}\left[k^{2}-
{\frac{a^{2}H^{2}}{2}(1-3\epsilon)}\right]+\mathcal{O}(\epsilon^{2}).
\end{equation}
By virtue of Eq.(\ref{11}), solving Eq.(\ref{8}) leads to
\begin{equation}\label{12}
\rho_{vac}=\frac{\Lambda_{c}^{4}}{16\pi^{2}}+(1+9\epsilon)\frac{H^{2}(t)\Lambda_{c}^{2}}{16\pi^{2}},
\end{equation}
\begin{equation}\label{13}
P_{vac}=\frac{\Lambda_{c}^{4}}{48\pi^{2}}-(1-3\epsilon)\frac{H^{2}(t)\Lambda_{c}^{2}}{48\pi^{2}}.
\end{equation}
The first term in Eqs.(\ref{12}) and  (\ref{13}) are the contribution of the energy density and pressure for Minkowskian space time; And because of the cutoff dependency the latter terms are well known bare quantities. To get rid of quartic divergencies the subtraction mechanism is a good suggestion, which is close to Casimir approach {\cite{10}}. The base of the Casimir effect is on the subtraction mechanism which the contribution of the energy for a Minkowski space-time and for example two plates which set in there are subtracted.  therefore when two infinite values for energy of space time and plates subtract one can obtain a finite quantity. Therefore by means of Arnowitt-Deser-Misner (ADM) approach \cite{{11}} and subtraction mechanism one concludes that the vacuum energy to an asymptotically flat space-time with metric $g_{\mu\nu}$ can be achieved as $E=H_{GR}(g_{\mu\nu})-H_{GR}(\eta_{\mu\nu})$, where $H_{GR}$  refers to the Hamiltonian  which is calculated in general relativity. This equation indicates that flat space-time does not gravitate and the contribution of the energy which is obtained in Minkowski space-time can be subtracted from related quantity in curved background \cite{{11}, {10}} and {\cite{11-a, 11-aa}}. Thence, because flat space time does not gravitate one able to subtract the contribution of quartic terms in Minkowski space from the same terms in Friedmann–Lemaitre–Robertson–Walker (FLRW) space time. Also it should be stressed the results which were obtained for de-Sitter like scenario have some notable differences with normal de-Sitter ones. The first is, for de Sitter like investigations the Hubble parameter is not  a constant and this causes to appearance of the first slow roll parameter in the model which could be considered to investigate the accuracy of this model. As second note, the coefficients which were appeared in energy density and pressure cause to increasing of the zero point quantum fluctuations contribution in the dark energy. Now let's investigate the bare quantities which are as
\begin{eqnarray}\label{14}
\rho_{bare}=(1+9\epsilon)\frac{H^{2}(t)\Lambda_{c}^{2}}{16\pi^{2}},
\end{eqnarray}
\begin{eqnarray}\label{14}
P_{bare}=-(1-3\epsilon)\frac{H^{2}(t)\Lambda_{c}^{2}}{48\pi^{2}},
\end{eqnarray}
 Following {\cite{12}} one can introduce counter terms for energy density and pressure respectively as
\begin{eqnarray}\label{14-aaa}
\rho_{count}&=&-(1+9\epsilon)\frac{H^{2}(t)\Lambda_{c}^{2}}{16\pi^{2}}+\rho_{Z}, \label{14-aa1}\\
P_{count}&=&(1-3\epsilon)\frac{H^{2}(t)\Lambda_{c}^{2}}{48\pi^{2}}+P_{Z},\label{14-aa2}
\end{eqnarray}
where
\begin{eqnarray}\label{14-aab}
\rho_{Z}&=&(1+9\epsilon)\frac{H^{2}(t)M^{2}}{16\pi^{2}},\label{15-a}\\
P_{Z}&=&-(1-3\epsilon)\frac{H^{2}(t)M^{2}}{48\pi^{2}}\label{15-b},
\end{eqnarray}
where subscript $Z$ refers to the zero point, $M$ is in order of Planck mass. It is notable in definitions of $\rho_{Z}$ and $P_{Z}$, both the Planck length, ultra violet cutoff, and Hubble length, infrared cutoff, are appeared which this result is in good agreement with observational results {\cite{akhma}}. Using  Eqs.(\ref{15-a}) and (\ref{15-b}), the equation of state (EoS) parameter for the vacuum fluctuations could be expressed as
\begin{equation}\label{16}
\omega_{Z}=\frac{P_{Z}}{\rho_{Z}}=\frac{{-1}}{{3}}+4\epsilon.
\end{equation}
 This relation indicates, the EoS of zero point quantum fluctuations is dependent on the first slow roll parameter. It should be noted also, that whereas this approach is similar to Casimir mechanism both positive and negative signs for energy density acceptable.  To consider this fact one can consider coefficient $\sigma=\pm1$, for Eq.(\ref{15-a}) and redefines it as
\begin{equation}\label{17}
\rho_{Z}=\sigma(1+9\epsilon)\frac{H^{2}(t)M^{2}}{16\pi^{2}}.
\end{equation}
 The positive sign causes an attractive force and negative ones is related to the repulsive case. Now for more discussions about time dependency of energy density of vacuum fluctuations, one able to redefine it based on critical energy density of the Universe. Hence considering  definition of critical energy density ($\rho_{cri}=3H^{2}(t)/8\pi G$) and by means of  definition of Planck mass $M_{Pl}={1}/\sqrt{G}$,  ($G$ is the Newtonian constant), the energy density of zero point fluctuations could be rewritten as
\begin{eqnarray}\label{18-b}
\rho_{Z}&=& \Omega_{Z} \rho_{cri}+\mathcal{O}(\epsilon^{2}), \cr
 &=& \beta \frac{1+9\epsilon}{\epsilon^{2}}a^{-2\epsilon}(t),
\end{eqnarray}
where $\Omega_{Z}={\beta (1+9\epsilon)}/{M^{2}_{Pl}}$, and $\beta=\sigma M^{2}/16 \pi^{2}$. Because of the time dependency of $\rho_{Z}(t)$, the conservation equation only for $\rho_{Z}$  does not satisfied. Hence  using $\dot{\rho}_{Z}=-2\epsilon H(t) \rho_{Z}$, and Eq.(\ref{18-b}) one has
\begin{equation}\label{19}
\dot{\rho}_{Z}+3H(t)\rho_{Z}(1+\omega_{Z})=Q,
\end{equation}
where $Q$ is dissipation function and it could be obtained as
\begin{equation}\label{20}
Q=2H(t) \rho_{Z}(1+5\epsilon).
\end{equation}
 Therefore the energy density of quantum fluctuations capable to  exchange energy with other components of the Universe. To investigate the transformation of energy we consider some different cases as follows.

 %=====================================================
%==============Investigation the  energy transfer  between different components of the Universe ====================
%=======================================================
  \section{Transformation of Energy  Between Different Components of the Universe}
%=====================================================
%==============Energy transfer  between zero point energy and  matter ====================
%=======================================================
 \subsection{Transformation of Energy  Between Zero Point Fluctuations and  Matter}
In this case one has
\begin{eqnarray}\label{21}
\dot{\rho}_{Z}+3H(t)\rho_{Z}(1+\omega_{Z})&=&2H(t) \rho_{Z}(1+5\epsilon),\cr
\dot{\rho}_{m}+3H(t)\rho_{m}&=&-2H(t) \rho_{Z}(1+5\epsilon),
\end{eqnarray}
 and therefore the combination two section of above equation yields
\begin{equation}\label{22}
\dot{\rho}_{Z}+3H(t)\rho_{Z}(1+\omega_{Z})+\dot{\rho}_{m}+3H(t)\rho_{m}=0,
\end{equation}
which indicates,  the conservation equation in general is satisfied. Therefore  by means of Eqs.(\ref{18-b}) and (\ref{22}), $\rho_{m}$ could be achieved as
\begin{equation}\label{23}
{\rho}_{m}={\Psi}{a^{-2\epsilon}(t)}+\tilde{\Psi}a^{-3}(t),
\end{equation}
where $\Psi= {-2\beta(1+44\epsilon/3)}/{3\epsilon^{2}}$ and $\tilde{\Psi}$ is integration constant. From Eq.(\ref{23}) it is realized that in our model the matter density equation is modified, where the first term indicates the matters which risen from interaction of quantum fluctuations with matter and the latter is indicated the remain contribution of  matter, namely ordinary cold dark matter.
%=====================================================
%==============Energy transfer  between zero point energy and  other contribution of DE====================
%=======================================================
 \subsection{Transformation of Energy Between Zero Point Energy and The Remanent Components of Dark Energy}
Assume there is an internal interaction between different components of dark energy, namely $\rho_{\Lambda}$ and $\rho_{Z}$ where $\rho_{\Lambda}$ indicates energy density of cosmological constant. In this case, one can suppose that $\omega_{\Lambda}=-1$ and therefore the conservation equation reads
\begin{equation}\label{24}
\dot{\rho}_{Z}+3H(t)\rho_{Z}(1+\omega_{Z})+\dot{\rho}_{\Lambda}=0.
\end{equation}
By virtue of $\dot{\rho}_{Z}={-2 \beta(1+3\epsilon)}/({\epsilon^2}a^{3\epsilon}(t))$ and definition of scale factor in  de-Sitter like background, one has
\begin{equation}\label{25}
{\rho}_{\Lambda}=\frac{\beta }{\epsilon}(1+14 \epsilon)H^{2}(t)+C_{0},
\end{equation}
where $C_{0}$ is integration constant.  In addition it is obvious that  because $\dot{{\rho}}_{Z}$  is  proportional to $\dot{{\rho}}_{\Lambda}$ the Big Bang Nucleosynthesis (BBN) constraint which has been discussed in {\cite{11}}, could be considered to estimate the upper  bound on $\Omega_{Z}={\beta (1+9\epsilon)}/{M^{2}_{Pl}}$. Also it should be stressed  by comparing ${\rho}_{Z}$ with one in normal de-Sitter model, it is clear that the coefficient $(1+9\epsilon)$ cause the increasing of the  magnitude of zero point energy density.
 %=====================================================
%==============Energy Transfer  between all components of the Universe====================
%=======================================================
\subsection{ Transformation of Energy  Between all Components of The Universe}
In this stage, one can consider a  general case which all components of the Universe are in an  interaction. Therefore the conservation equations could be written as
\begin{eqnarray}
\dot{\rho}_{m}+3H(t)\rho_{m}&=&\tilde{Q}_{i}, \label{27-a}
\end{eqnarray}
\begin{eqnarray}
\dot{\rho}_{\Lambda}+\dot{\rho}_{Z}+3H(t)\rho_{Z}(1+\omega_{Z})&=&-\tilde{Q}_{i}, \label{27-b}
\end{eqnarray}
where $\tilde{Q}_{i}$ are dissipation functions and are defined as follows
\begin{itemize}
\item . $\tilde{Q}_{1}=3 \kappa H(t)\rho_{\Lambda}$,
\item . $\tilde{Q}_{2}=3 \gamma H(t) \rho_{m}$,
\item . $\tilde{Q}_{3}=3 \theta H(t) \rho_{Z}$,
\end{itemize}
and also $\kappa$, $\gamma $ and $\theta$ indicate the strength of the interaction between different components of the Universe {\cite{15}}.
 %=====================================================
%==============Solving conservation equation for $\tilde{Q}_{1}$====================
%=======================================================
\subsubsection{Solving Conservation Equation for $\tilde{Q}_{1}$ }
by virtue of Eq.(\ref{27-b}) and dissipation function $\tilde{Q}_{1}$, the conservation equation for dark energy components of the Universe is as
\begin{eqnarray}\label{28}
\dot{\rho}_{\Lambda}+\dot{\rho}_{Z}+3 \kappa H(t)\rho_{\Lambda} +3H(t)\rho_{Z}(1+\omega_{Z})=0,
\end{eqnarray}
based on Eqs.(\ref{18-b}) and (\ref{16}), then Eq.(\ref{28}) could be rewritten as
\begin{eqnarray}\label{29}
\dot{\rho}_{\Lambda}+3 \kappa H(t)\rho_{\Lambda} +{2\beta(1+14\epsilon) H^{3}(t)}=0,
\end{eqnarray}
hence solving this differential equation  yields
\begin{eqnarray}\label{30}
{\rho}_{\Lambda}=\frac{-\breve{D}}{(3{\kappa}-2 {\epsilon})}a^{-2\epsilon}(t)+\tilde{c}a^{-3\kappa},
\end{eqnarray}
where $\breve{D}=2 (1+14\epsilon)\beta/\epsilon^{2}$ and $\tilde{c}$ is integration constant. By substituting Eq.(\ref{30}) into   (\ref{27-a}) one can attain $\rho_{m}$ as follows
\begin{eqnarray}\label{31}
{\rho}_{m}={\frac{{-\breve{D}}}{\Big(3-2{\epsilon}(1-\frac{1}{\kappa})\Big)}}a^{-2\epsilon}(t)+\frac{\kappa \tilde{c}}{1-\kappa}a^{-3\kappa}+\breve{B}a^{-3}(t).
\end{eqnarray}
In above equation $\breve{B}$ is integration constant. From this relation one can conclude that,  in matter equation only ordinary cold dark matter does not appeared, rather an extra term is appeared which is risen from interaction of matter and quantum fluctuations, namely sub-dark matter. In the following, we will come back to this issue.
 %=====================================================
%==============Solving conservation equation for $\tilde{Q}_{2}$ ====================
%=======================================================
\subsubsection{Solving Conservation Equation for $\tilde{Q}_{2}$ }
By rewriting Eq.(\ref{27-a}) and (\ref{27-b})  for   $\tilde{Q}_{2}$, one arrives
\begin{eqnarray}
\dot{\rho}_{m}+3H(t)\rho_{m}=3 \gamma  H(t) \rho_{m},\label{32}
\end{eqnarray}
\begin{eqnarray}
\dot{\rho}_{\Lambda}+\dot{\rho}_{Z}+3H(t)\rho_{Z}(1+\omega_{Z})=-3 \gamma  H(t) \rho_{m},\label{32-b}
\end{eqnarray}
thus, Eq.(\ref{32}), could be considered as
$$\dot{\rho}_{m}+3H(t)(1-\gamma )\rho_{m}=0,$$
and after solving, one can attain
\begin{eqnarray}\label{34}
\rho_{m}= \rho_{m0}a^{-3(1-\gamma )}(t).
\end{eqnarray}
Substituting Eq.(\ref{34}) into Eq.(\ref{32-b}), yields
\begin{eqnarray}\label{35}
{\rho}_{\Lambda}=\frac{\beta(1+14\epsilon)}{\epsilon}H^{2}(t)+\frac{\rho_{m0}\gamma }{3(1-\gamma )}a^{-3(1-\gamma )}+\mathfrak{M},
\end{eqnarray}
where $\mathfrak{M}$ is the integration  constant.
 %=====================================================
%==============Solving conservation equation for $\tilde{Q}_{3}$  ====================
%=======================================================
\subsubsection{Solving Conservation Equation for $\tilde{Q}_{3}$ }
In this stage, one can suppose that the interaction between different components of the Universe is determined by virtue of $\tilde{Q}_{3}$. Thence Eqs.(\ref{27-a}) and (\ref{27-b}), are rearranged as
\begin{eqnarray}\label{37}
\dot{\rho}_{m}+3H(t)\rho_{m}&=&3 \theta H(t) \rho_{Z},
\end{eqnarray}
\begin{eqnarray}\label{38}
\dot{\rho}_{\Lambda}+\dot{\rho}_{Z}+3H(t)\rho_{Z}(1+\omega_{Z})&=&-3 \theta H(t) \rho_{Z}.
\end{eqnarray}
Using definition of scale factor in de-Sitter like background, one allows to rewrite Eq.(\ref{37}) as
\begin{eqnarray}\label{39}
\dot{\rho}_{m}+\frac{3}{\epsilon}{\rho}_{m} a^{-\epsilon}-\frac{3\theta}{\epsilon}\bigg(\frac{\beta(1+3\epsilon)}{\epsilon^2}\bigg)a^{-3\epsilon}=0,
\end{eqnarray}
solving this equation for ${\rho}_{m}$, yields
\begin{eqnarray}\label{40}
{\rho}_{m}=\breve{A}a^{-2\epsilon}+\breve{K}a^{-3}(t),
\end{eqnarray}
where $\breve{A}\simeq{3\theta}{\beta(1+8\epsilon)}/{\epsilon^2}$ and $\breve{K}$ is the integration constant.
Hence, to attain ${\rho}_{\Lambda}$, from Eq.(\ref{38}) one has
\begin{eqnarray}\label{41}
\dot{\rho}_{\Lambda}+\frac{\beta}{\epsilon^3}\big[2+3\theta+27\epsilon(1+\theta)\big]a^{-3\epsilon}(t)=0,
\end{eqnarray}
solution of this equation for ${\rho}_{\Lambda}$, reads
\begin{eqnarray}\label{42}
{\rho}_{\Lambda}=\frac{3\beta}{\epsilon}\big[2+3\theta+27\epsilon(1+\theta)\big]H^{2}(t).
\end{eqnarray}
%=====================================================================
%======================= Bounds which risen from time evolution of DE======================
%=====================================================================
\section{Bounds which risen from time evolution of dark energy}
\begin{figure}
\begin{center}
\includegraphics[width=5cm]{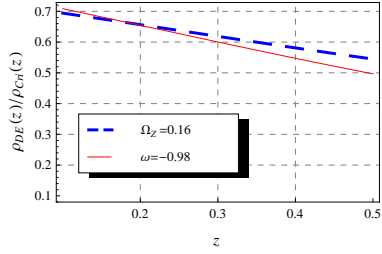}
\caption{\label{Fig.1}{\small ${\rho_{DE}(z)}/{\rho_{cri}(z)}$  versus $z$ have been shown with dashed-blue line, Eq.(\ref{48}), and solid-red line respectively, Eq.(\ref{50-a}). The auxiliary parameters are $\Omega_{Z} = 0.16$, $\Omega_{m} = 0.17$, $\tilde{\Omega}_{m}=0.1$, $\Omega_{DE} = 0.73$, $\epsilon= 0.02$  and $\omega_{\Lambda}= -0.98$.}}
\end{center}
\end{figure}

\begin{figure}
\begin{center}
\includegraphics[width=5cm]{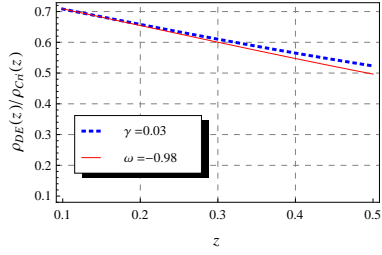}
\caption{\label{Fig.2}{\small ${\rho_{DE}(z)}/{\rho_{cri}(z)}$   versus $z$   have been shown with dotted line, Eq.(\ref{50}), and  solid line,  Eq.(\ref{50-a}). The auxiliary parameters are  $\Omega_{Z} = 0.16$, $\Omega_{m} = 0.23$, $\Omega_{DE} = 0.73$, $\gamma = 0.03$  and $\omega_{\Lambda}= -0.98$.}}
\end{center}
\end{figure}

\begin{figure}
\begin{center}
\includegraphics[width=5cm]{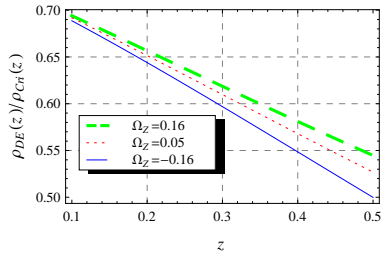}
\caption{\label{Fig.3}{\small ${\rho_{DE}(z)}/{\rho_{cri}(z)}$ versus $z$,  Eq.(\ref{48}), for three different values  $\Omega_{Z} = 0.16$ , $\Omega_{Z} = 0.05$ and $\Omega_{Z} = -0.16$. They have been shown with dashed-green , dotted-red  and blue (solid) lines respectively}}.
\end{center}
\end{figure}

In this section we want to compare the results of this work with results which risen from standard $\Lambda$CDM model. For this end, one can start from the Friedmann equation.  Therefore the ratio of dark energy density and critical energy density in the standard model as a function of red shift parameter, $z$, is obtained as
\begin{eqnarray}\label{50-a}
\frac{\rho_{DE}(t)}{\rho_{cri}(t)}=\frac{\Omega_{\Lambda}(1+z)^{3(1+\omega_{\Lambda})}}{\Omega_{M}(1+z)^{3}+\Omega_{\Lambda}}.
\end{eqnarray}
It should be emphasized in above equation, one can get  $\Omega_{\Lambda}=0.73$,  $\Omega_{M}=0.23$ and $\omega_{\Lambda}=-0.98$, which are obtained from a combination of CMB, BAO and SNeIa data sets \cite{ {11}, {1N}, {2N}, {16}}. Whereas in this work, the components of dark energy are as quantum fluctuations and cosmological constant, thence the Friedmann equation is obtained as
\begin{eqnarray}\label{43}
\frac{3H^{2}(t)}{8\pi G}&=&\bigg(\rho_{m}(t)+\rho_{DE}\bigg),\cr
&=& \bigg(\rho_m(t)+\rho_\Lambda(t)+\Omega_{Z}\rho_{cri}(t)\bigg),
\end{eqnarray}
where subscript DE indicates dark energy. By substituting Eq.(\ref{18-b}) into Eq.(\ref{43}), the Friedmann equation could be rewritten as
\begin{eqnarray}\label{44}
H^{2}(t)=\frac{H_{0}^{2}(t)}{1-\Omega_{z}}\bigg(\Omega_{m}(t)+\Omega_{\Lambda}(t)\bigg).
\end{eqnarray}
Where $\Omega_{i}(t)=\rho_{i}(t)/\rho_{cri}(0)$, and $i$ refers to $m$, $\Lambda$ and $Z$ respectively; In addition $\rho_{cri}(0)$ denotes the critical energy density in present epoch. It should be noted the  energy densities of curvature and radiation are neglected. From Eq.(\ref{18-b}) and definition of dimensionless energy density parameters, one gets
\begin{eqnarray}\label{45}
\rho_{DE}=\Omega_{Z}\rho_{cri}(t)+\rho_{cri}(0)\Omega_{\Lambda}.
\end{eqnarray}
By virtue of definition of $\rho_{cri}(t)$, the Eq.(\ref{44}) could be rearranged as
\begin{eqnarray}\label{46}
\rho_{cri}(t)=\frac{\rho_{cri}(0)}{1-\Omega_{Z}} \bigg[\Omega_{m}(t)+\Omega_{\Lambda}\bigg].
\end{eqnarray}
By dividing $\rho_{DE}$ and $\rho_{cri}(t)$, one arrives
\begin{eqnarray}\label{47}
\frac{\rho_{DE}(t)}{\rho_{cri}(t)}=\Omega_{Z}+\frac{(1-\Omega_{Z})\Omega_{\Lambda}}{\Omega_{m}(t)+\Omega_{\Lambda}}.
\end{eqnarray}
Whereas, different equations for $\rho_{m}$ (based on different conditions for interaction) are attained,  $\Omega_{m}(t)$ in Eq.(\ref{46}) gets different forms. Thus by considering Eqs.(\ref{23}), (\ref{31}), (\ref{40}) and Eq.(\ref{47}) it could be rewritten respectively as
\begin{itemize}
\item[a.] From Eq.(\ref{23}):
\begin{eqnarray}\label{48}
\frac{\rho_{DE}(t)}{\rho_{cri}(t)}=\Omega_{Z}+\frac{(1-\Omega_{Z})(\Omega_{DE}-\Omega_{Z})}{\Omega_{m}(1+z)^{3}+
\tilde{\Omega}_{m}(1+z)^{2\epsilon}+(\Omega_{DE}-\Omega_{Z})},
\end{eqnarray}
where $\tilde{\Omega}_{m}={\Psi}/\rho_{cri}(0)$ and ${\Omega}_{m}=\tilde{\Psi}/\rho_{cri}(0)$. In Figure \ref{Fig.1}, the evolution of  Eq.(\ref{48}) versus $z$  parameter shows a deviation in comparison to the ordinary ${\rho_{DE}(t)}/{\rho_{cri}(t)},$ Eq.(\ref{50-a}), that illuminates the effects of $\Omega_{Z}$ and $\epsilon$ in the evolution of this function.
%%%%%%%%%%%%%%%%%%%%%%%%%%%
%%%%%%%%%%%%%%%%%%%%%%%%%%
\item[b.] From Eq.(\ref{31}):
\begin{eqnarray}\label{49}
\frac{\rho_{DE}(t)}{\rho_{cri}(t)}=\Omega_{Z}+\frac{(1-\Omega_{Z})(\Omega_{DE}-\Omega_{Z})}{\Omega_{m}(1+z)^{3}+
\bar{\Omega}_{m}(1+z)^{2\epsilon}+{\Omega}^{\ast}_{m}(1+z)^{3\kappa}+(\Omega_{DE}-\Omega_{Z})},
\end{eqnarray}
where $\bar{\Omega}_{m}={{-\breve{D}}}/{\rho_{cri}(0)\big(3-2{\epsilon}(1-1/\kappa)\big)}$ and ${\Omega}^{\ast}_{m}={{\kappa}\tilde{c}}/{{\rho_{cri}(0)}(1-\kappa)}$. Considering the above equation, it is find out the $\tilde{Q}_{1}$ case, does not lead to physical result, because the equation for $\rho_{m}$ can not satisfy observations as well.
\item[c.] From Eq.(\ref{34})
\begin{eqnarray}\label{50}
\frac{\rho_{DE}(t)}{\rho_{cri}(t)}=\Omega_{Z}+\frac{(1-\Omega_{Z})(\Omega_{DE}-\Omega_{Z})}{\Omega_{m}(1+z)^{3(1-\gamma )}
+(\Omega_{DE}-\Omega_{Z})}.
\end{eqnarray}
where $\Omega_{m}(1+z)^{3(1-\gamma )}$  term, indicates dark matter component of the Universe.
Figure \ref{Fig.2} shows a deviation with respect to ordinary ${\rho_{DE}(t)}/{\rho_{cri}(t)}, $ Eq.(\ref{50-a}), which in comparison to Figure \ref{Fig.1} it is realized that $Q_{2}$ function has better results.
%%%%%%%%%%%%%%%%%%
%%%%%%%%%%%%%%%%%
\item[d.] From Eq.(\ref{40}):
\begin{eqnarray}\label{51}
\frac{\rho_{DE}(t)}{\rho_{cri}(t)}=\Omega_{Z}+\frac{(1-\Omega_{Z})(\Omega_{DE}-\Omega_{Z})}{\Omega_{m}(1+z)^{3}+{\Omega}^{\sharp}_{m}(1+z)^{2\epsilon}+(\Omega_{DE}-\Omega_{Z})}.
\end{eqnarray}
where $\Omega_{m}=\breve{K}/{\rho_{cri}(0)}$ and ${\Omega}^{\sharp}_{m}=\breve{A}/{\rho_{cri}(0)}$. The energy density parameter in this case, behaves as the case related to Eq.(\ref{25}). This manner is similar to case $a$  i.e. Eq.(\ref{48}), and therefore, it does not need to plot it's evolution. It should be noted that  the behaviour of ${\rho_{DE}(z)}/{\rho_{cri}(z)}$  based on different quantities  for $\Omega_{Z}$  are plotted in Figure \ref{Fig.3}.
\end{itemize}
As it is obvious, Figure \ref{Fig.3} illuminates the evolution of ${\rho_{DE}(z)}/{\rho_{cri}(z)}$ versus $z$,  Eq.(\ref{48}), for three different values  $\Omega_{Z} = 0.16, 0.05$ and $-0.16$. The results  are shown with  green dashed, red dotted and blue solid  lines respectively. It is clear that ${\rho_{DE}(z)}/{\rho_{cri}(z)}$  decreases with decreasing  $\Omega_{Z}$.
%=====================================================================
%======================= Conclusion and discussion======================
%=====================================================================
\section{Conclusion and discussion}
Vacuum quantum fluctuations  in a de-Sitter like background for both massive and massless bosonic scalar field have been investigated. In light of Planck database $2013$ we have estimated the best value for the first slow roll parameter and using such quantity, the different components of the Universe's energy budget have been calculated. It should be stressed, the scalar fields have quantized in a FLRW framework and it have  shown that the contribution of vacuum fluctuations have increased in such  background in comparison with normal de-Sitter case. It should be emphasized, the subtraction approach have been used to eliminate the infinities which were appeared in the calculations. Incidentally, using the physical energy density of zero point quantum fluctuation, it has been realized that this component of the Universe have to had an interaction with other components of the Universe. In addition, when the energy density of matter are achieved, it has been found, that beside of ordinary dark matter there are exist components of matter which were created due to interaction with zero point quantum fluctuations. Also whereas zero point energy density was  time dependent, the transformation of enenrgy between different ingredients of the Universe have been investigated. It was considerable that, for the state in which all components of the Universe exchange energy between themselves, time evolution of  ${\rho_{DE}(z)}/{\rho_{cri}(z)}$ have been shown that $Q_{2}=3 \gamma H \rho_{m}$ is in best agrement in comparison to observational database and also the interaction term as $\tilde{Q}_{1}=3 \kappa H(t)\rho_{\Lambda}$, had not any physical results. At last for more details, the bounds which have risen from time evolution of dark energy  density in comparison to standard $\Lambda$ cosmology have been investigated.  To compare the results of this work with observational data, we have regarded  the time evolution of  ${\rho_{DE}(z)}/{\rho_{cri}(z)}$ which concluded from a combination of CMB, BAO and SNeIa data sets. From Figures \ref{Fig.2} and \ref{Fig.3}, the evolution of Eqs.(\ref{48}) and (\ref{50}) versus $z$ in comparison to observational results have been illustrated.

\section*{Acknowledgment}
HS would like to thank Iran's National Elites Foundation for financially support during this work. He  expresses his appreciation to the Prof. Y. Sobouti for sharing their pearls of wisdom with him during the course of this research.
%=====================================================================
%======================= Reference ===================================
%=====================================================================

\end{document}